# Study of Security Issues in Pervasive Environment of Next Generation Internet of Things

Tapalina Bhattasali[1], Rituparna Chaki[2], and Nabendu Chaki[3]

[1,3] Department of Computer Science & Engineering
University of Calcutta, Kolkata, India
[1]`tapolinab@gmail.com`, [3]`nabendu@ieee.org`
[2] A.K. Choudhury School of IT
University of Calcutta, Kolkata, India
[2]`rituchaki@gmail.com`

**Abstract.** Internet of Things is a novel concept that semantically implies a world-wide network of uniquely addressable interconnected smart objects. It is aimed at establishing any paradigm in computing. This environment is one where the boundary between virtual and physical world is eliminated. As the network gets loaded with hitherto unknown applications, security threats also become rampant. Current security solutions fail as new threats appear to destruct the reliability of information. The network has to be transformed to IPv6 enabled network to address huge number of smart objects. Thus new addressing schemes come up with new attacks. Real time analysis of information from the heterogeneous smart objects needs use of cloud services. This can fall prey to cloud specific security threats. Therefore need arises for a review of security threats for a new area having huge demand. Here a study of security issues in this domain is briefly presented.

## 1 Introduction

Kevin Ashton, cofounder and executive director of MIT Auto-ID Centre, first introduced the term "Internet of Things" in 1999 [1]. It is an emerging concept where smart objects are equipped with sensors or actuators, tiny microprocessor, communication interface, and power resource. As per Libelium report [2], it is estimated that more than 50 billion devices will be accessing the Internet by the year 2020. The tremendous increase in demand makes mapping of resource constrained sensor nodes and Internet a big challenge. The term 'Thing' used in "Internet of Things" can be defined as physical or digital or virtual entity that is capable of being identified [3] to integrate heterogeneous data, semantics, and objects. IoT combines several techniques such as RFID,Zigbee,Wi-Fi,3G/4G,embedded devices, sensing devices.

To ensure sensing data availability at any time, at any place, effective processing of large amounts of collected sensing data is necessary in the application areas such as environmental monitoring, weather forecasting, transportation, business, healthcare, military application etc. Combining wireless sensor networks with cloud makes it easy to share and analyze real time sensor data on-the-fly. The issues of storage capacity may be overcome by low-cost cloud computing technique [4][5]. For security and easy access of data, it is widely used in distributed and mobile environment. The objective of the integrated sensor-cloud framework is to facilitate the shifting of high volume of sensing data from sensor networks to the cloud computing environment; so that scientifically and economically feasible data can be fully utilized to visualize the concept of next generation Internet i.e. Internet of Things. Figure 1 represents "any" paradigm in the context of IoT. In order to provide anytime, anywhere services, a number of smart sensing objects attached to Internet have to communicate with each other through uniquely assigned identity [6]. This is where128 bits IPv6 steps in. So that, it can support $2^{128}$ addresses, which is approximately 340 undecillion or $3.4 \cdot 10^{38}$ addresses.

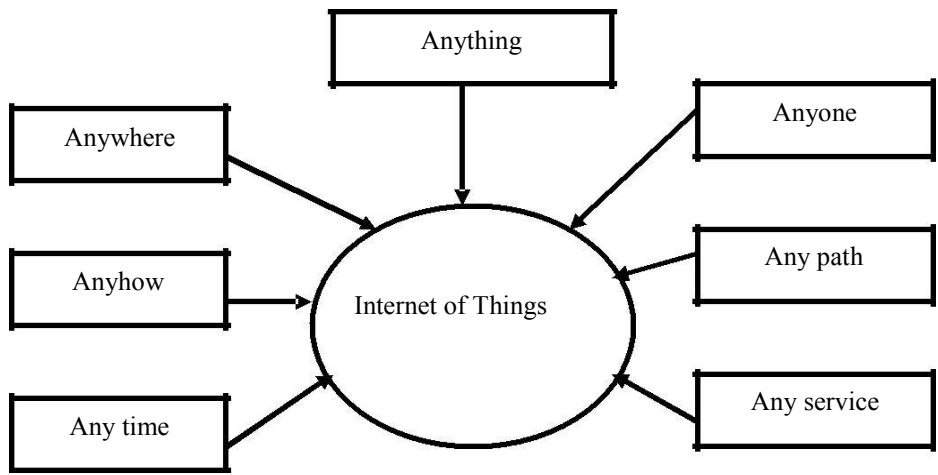

**Fig. 1.** "Any" Paradigm in IoT

The design of protocol stack for smart objects must be matched with existing Internet hosts in order to create the extended Internet, which is the aggregation of Internet with the IoT. It is very critical to implement IPv6 in low powered sensing objects. As IPv4 address spaces are already used, existing IPv4 (32 bits) enabled systems need to be up-graded to IPv6 compatible devices. The huge number of heterogeneous devices being connected to one another gives rise to newer security threats. The amount of flexibility and availability of information also leads to hitherto unforeseen security breaches. At present, there is a lack of analyzers for the new threats in the network. Securing IP based ubiquitous sensor network (IPv6 or IPv4-Ipv6) in IoT is of great concern for researchers to meet future market demands and satisfaction. In comparison to IPv4, IPv6 provides simplicity, improved routing speed, quality of service and security. IPv6 brings signifi-cant assurance of a higher level of security and confidentiality of the transmitted data.

The heterogeneous, resource-constrained and distributed natures of the network make conventional security methodologies inefficient. More research works are required to ensure security, performance and interoperability between next generation IoT and ex-isting Internet.

The remainder of this paper is organized as follows. Section 2 introduces the concept of next generation network i.e. IoT. Section 3 illustrates open issues in IoT environment. In section 4, possible security threats are explained briefly. Section 5 gives basic idea about security solutions in IoT environment. Section 6 discusses about a case study in this environment. It is followed by a conclusion in section 7.

## 2      Next Generation Internet of Things

IoT is an integrated part of future Internet, which may be termed as next generation network. Ambient intelligence which is hidden in IoT environment supports people in carrying out their everyday activities in an easy and natural way. Radio-frequency identification (RFID) is often seen as essential requirement for serving Internet of Things. If all objects around the environment are equipped with radio tags, they can be easily identified. This technology is very useful in health monitoring applications such as automated monitoring of patient's heart condition. In order to track patients' medical history, RFID chips can be implanted in their body. Internet-connected devices can also directly communicate with the required emergency services when sensed data shows the sign of deterioration in the patient's condition .In this way pervasive environment of smart healthcare technique has the capacity to save lives.

Pervasive computing, which is often similar to ubiquitous computing, is an important field of research leading to the domain of IoT. Tremendous developments in technologies such as wireless communications, mobile computing, wearable computers, sensors, RFID tags have led to the evolution of next generation IoT. The goal of pervasive computing is to create ambient intelligence where devices embedded in the environment always provide services to improve quality of life by hiding underlying technologies. In this pervasive environment, intelligent objects are interconnected to autonomously collect, process and transport data in a cooperative way, in order to adapt IoT concept [7]. It can be said that, IoT is more vulnerable to serious security threats than existing network [1] because it includes various constraint such as low resource objects, heterogeneous nature, open environment deployment, dynamic nature. In the next section, some of the considerable open issues for security implementation in IoT are discussed.

## 3      Open Issues in IoT Environment

In IoT environment, the system needs a flexible security mechanism to update it easily according to the requirement. Both the user and the system need a secure access control and authorization mechanism within the limitations imposed by IoT environment. Some of the open issues are as follows.

- IoT creates a heterogeneous environment facilitated by mobility of the objects. It can give rise to inconsistent interpretations of data collected from different domains. Due to distributed and ad-hoc nature, it is open to several unique vulnerabilities whose solutions are unknown. There is no central control that can provide required security features. Therefore burden of the security features may be too large for small and limited capacity objects.
- In this surrounding, a secure communication channel is needed along with object authentication to track interacting objects. But request for establishing secure channel is also transmitting through the shared, unreliable wireless medium. Therefore feasible solution for tracking objects through insecure channel is still in the phase of research.
- If data is shared with unknown objects, the probability of data security reduces automatically. The exact criteria for trust establishment between communicating parties need to be determined.
- To work in a smarter way, pervasive environment needs to deal with users' personal data. But sometimes it poses serious threat to the privacy of the user, especially in the situations where people do not want to disclose their detailed personal information.

Therefore an intelligent system should be considered to analyze these issues and to adapt dynamic mechanism. Next section focuses on possible security threats in IoT environment.

## 4    Possible Security Threats

Various security challenges can arise in heterogeneous features of IoT [8]. There are several known attacks in IoT environment whose solutions are available in the market. But aim of this paper is to consider novel challenges in IoT environment whose feasible solutions are still under research. Most of the security issues such as eavesdropping, false routing, message tampering, unauthorized usage, DoS attack are common with existing internet. But the issues related to specific attacks may be quite different. Some issues like secret extractions, tampering of nodes are more serious in IoT environment. This paper considers that security challenges in IoT may occur from compromised wireless sensor network, usage of novel IPv6 protocol, integrated cloud environment, smart sensing objects. In our previous papers [9][10][11], we have already discussed about security related issues in wireless sensor network. Now the main focus is to identify new threats in IPv6 based integrated sensor-cloud environment, which is based on number of smart sensing objects.

### 4.1    Security Attacks Related to IP-Enabled Environment

There are several existing threats in IP-enabled environment. Security in IPv6 is almost same as IPv4 security in many ways. IPv6 is normally considered as more secure than IPv4 because version 6 includes the concept of IP Security (IPSec). Beside this, some significant differences exist between IPv4 and IPv6 [12]. In the transition period, coexistence of IPv4 and IPv6 especially creates problem regarding security issues. It is because transition mechanisms provide new, previously unknown possibilities of intrusion. There are several transition mechanisms, such as tunneling, dual-stack configurations. It is very important to understand security implications of the transition mechanisms in order to apply proper security mechanisms. On dual-stack configuration hosts applications can be targeted by both IPv4 and IPv6 attacks. Tunneling mechanisms may also bring misuse possibilities. Tunneling, especially automatic tunneling can facilitate an intruder to avoid ingress filtering checks. Among two major methods of automatic tunneling, "6 to 4" method encapsulates IPv6 packet directly into an IPv4 packet and "Teredo" method encapsulates IPv6 packet into an IPv4 UDP packet. By misusing 6to4 transition mechanism, a DoS attack can be targeted to IPv6 node, IPv4 node or other 6 to 4 node. In 'Teredo" tunneling, all receiving nodes must allow decapsulation of packets that can be sourced from anywhere. This can be a serious security problem. Addresses within IPv4 and IPv6 headers may be spoofed to be used for Denial of Service (DoS) attacks. In this section, the main focus is on new attacks which may arise in IPv6 or IPv6-IPv4 environment. Some of the specific attacks in IP enabled network are briefly discussed.

☐ Reconnaissance Attack

IPv4 reconnaissance attack uses ping sweep and port scan techniques. It can be mitigated by filtering certain types of messages used by an intruder. The default subnet size of an IPv6 subnet is 64 bits, or $2^{64}$, compared to the subnet size in IPv4 of 8 bits, or $2^8$. This increases the scan size to check each host on a subnet by $2^{64} - 2^8$ (approximately 18 quintillion). So ping sweep and port scan are much more difficult to complete in an IPv6 network. New multicast addresses in IPv6 enable intruder to identify key resources such as routers more easily. Additionally, IPv6 networks are even more dependent on ICMPv6 to function properly. Aggressive filtering of ICMPv6 can have negative effects on network functions. As public services on the network need to be reachable with DNS, adversary can attack at least a small number of critical hosts within the victim network. DNS Server can be easily compromised because of using dynamic DNS mechanisms for large nature of IPv6 addresses and the lack of a strict requirement for Network Address Translation (NAT).

☐ Fragmentation Attack

Minimum recommended MTU size for IPv6 is 1280 octets. IPv6 protocol specification does not allow packet fragmentation by intermediary devices. It is recommended to drop all fragments less than 1280 octets unless the packet is the last in the flow. However an intruder may achieve port numbers by using fragmentation and can bypass security monitoring devices. An attacker can also cause overload of reconstruction buffers on the target system by sending a large number of small fragments, which forces a system to crash.

☐ ICMPv6 Misuse Attack

Some important mechanisms in IPv6 networks, such as neighbor discovery and path MTU discovery, are dependent on some types of ICMPv6 messages. ICMPv6 allows error notification response to be sent to multicast addresses, which can be misused by attacker. An attacker can cause multiple responses targeted at the victim by sending packet to a multicast address.

- Routing Header Misuse Attack

IP options in IPv4 are replaced with extension headers in IPv6. All IPv6 nodes are capable to process routing headers. It is possible that an intruder sends a packet to a publicly accessible address with a routing header containing forbidden address i.e. address of the victim network. Then the publicly accessible host will forward the packet to a destination address stated in the routing header even though that destination address is filtered.

There are several known and unknown security threats which can arise in IP enabled network. Some of the devastating attacks are discussed here. Next subsection describes security threats of the sensor-cloud environment.

### 4.2 Security Threats of Sensor-Cloud Environment

Sensor-cloud environment can be easily compromised by the adversary because of the absence of centralized control[13]. In this virtual environment, main security issue includes violation of authentication and leakage in communication channel. As sensor-cloud framework is deployed in distributed environment, there may appear several security challenges [14] [15] [16] that comes from the following perspectives. Environment of sensor nodes can be compromised or individual sensor can be vulnerable to attacks, whose solutions are available. Information flows within the cloud can be affected by compromised cloud nodes. The cloud client can be infected by malicious code, which can lead to further security breaches within the environment. The communication channels between sensors and cloud and between client and cloud are vulnerable to different types of attacks. Some of the typical attacks in this environment are SPAM over Internet Telephony (SPIT), where an attacker sends bulk unsolicited calls to an enterprise; The spammer attempts to initiate a voice session and then relays a prerecorded message if the callee answers; Denial of Service (DoS), which is an severe issue for any IP network- based service; Service theft, where an unauthorized users get access to the network that results into authentication violation.

In the next subsection, the main focus is on security risks for the smart objects in IoT.

### 4.3 Security Risks for Smart Objects Which Form Basis of IoT

In IoT environment, the solutions of some known attacks such as man-in-the-middle attack, eavesdropping, routing attacks are already known. The transmission phase in IoT environment may be vulnerable to man-in-the middle attacks, where it is assumed that no third party is able to sit in between the two communicating entities. In completely automated mechanisms, there is usually no prior knowledge about each other and can not always be able to identify intruder. During transmission between smart objects in a network, it may be susceptible to eavesdropping, either for insufficient protection of communication medium or for use of compromised session key. Routing information in IoT can be spoofed, altered, or replayed. Other known relevant routing attacks include sinkhole attack or blackhole attack, selective forwarding, wormhole attack, sybil attack. Other security threats related to smart sensing objects are as follows [17][8].

□ Privacy Threat

Tracking of object's location and usage may give rise to privacy risk for its users. An attacker can gather information about individual object to find out behavioral patterns of the users. Such information can be distributed to interested parties for marketing purposes. A smart object deployed in the ambient environment can easily be captured by an attacker. Such an attacker may then attempt to extract security information to misuse it.

□ Firmware Replacement Attack

When a smart object is in operation or maintenance phase, its firmware or software may be updated to allow for new functionality or new features. An attacker may be able to exploit such a firmware upgrade by replacing the object with malicious software, thereby influencing the operational behavior of the object.

□ Cloning of Smart Objects by Untrusted Manufacturer

During the manufacturing process of a smart object, an untrusted manufacturer can easily clone the physical characteristics or security configuration of the object. Cloned object may be sold at a cheaper price in the market, and be able to function normally as a genuine object. In the worst case, a cloned object can be used to control a genuine object. An untrusted manufacturer may also change functionality of the cloned object, resulting in degraded functionality with respect to the genuine object. It can also implement additional functionality with the cloned object, such as a backdoor. During the installation of object, a genuine object may be substituted with a similar variant of lower quality without being detected. The main motivation may be cost savings. Genuine objects can be resold in order to gain further financial benefits. Another motivation may be to damage reputation of competitors.

After going through different security threats from heterogeneous perspective, next section focuses on possible security solutions in IoT.

## 5   Security Solutions in IoT Environment

Security in IoT environment should address the following main issues.

- □ Enabling smart and intelligent behavior of networked objects.
- □ Preservation of privacy for heterogeneous sets of objects.
- □ Decentralised authentication and trust model.
- □ Energy efficient security solutions.
- □ Proper authentication of the objects within the network.
- □ Security and trust for cloud computing services.
- □ Data ownership.

Security should ensure accurate implementation of confidentiality, integrity, authentication, non-repudiation and access control. Two of the main security issues in the IoT are privacy and confidentiality. Because of the scale of deployment and mobility, the cloud of "things" is hard to control. Cryptographic techniques are useful to protect confidential information stored in the network and to transfer secure messages from one ubiquitous node to another. But there are some gaps between available techniques and requirements. Implementing cryptography into the network of smart objects creates tremendous challenges. Because traditional mechanisms are slow in speed, large in size and consume more power and may fail to provide necessary protection for sensed data. Lightweight Cryptography[18][19] is a current field of research that can meet the constraints set by the use of smart objects. There is no strict criterion for classifying a cryptographic algorithm as lightweight but the basic concept is that cryptography techniques need to work by using minimum amount of essential resources of target objects. It can be categorized into hardware-oriented and software-oriented. Hardware-oriented techniques are more applicable in areas where main concern is about the size of chip and number of clock cycles required for its execution. Software oriented techniques need to be considered when main focus is on memory (Ram and ROM) requirements, and power consumption. Another type of categorization is symmetric versus asymmetric. Asymmetric cryptographic algorithms offer more security than symmetric, but symmetric technique works well, where authentication and integrity are of prime importance than non-repudiation and confidentiality. This can save additional computational cost and power consumption. Asymmetric ciphers are computationally far more demanding in both hardware and software levels. Finally it is the designer's decision to choose the appropriate techniques based upon the application's requirements along with the constraints carried by them.

In some cases, maximum security can only be achieved by designing an effective intrusion detection system which is not intended to prevent attacks. Instead, their purpose is to provide alert about possible attacks, ideally in time to stop the attack or to mitigate the damage. Researchers have been working for quite some time for designing intrusion detection in wireless sensor network and in IP based network. IoT involves heterogeneous network and thus an integration of techniques is needed. Not much work has been done in the field of IoT environment. There exist some open source IDS systems for IPv4 network. By using software, IDS systems in IPv4 networks, procedure of intrusion detection can be automated. In this, intrusion attempt is recognized and logged by IDS system and alert is generated. There is no freeware IDS software for IPv6 networks. By using packet analyzer tools, an intrusion detection procedure will require an efficient network administrator. Therefore IPv6 supporting IDS system must consider IPv6 protocol specific features.

- IPv6 defines a new header format, which must be properly recognized by IDS.
- IDS must implement support for IPv6 extension headers and to check the order of extension headers. It is recommended for IDS to discard a packet with an undefined "Next Header" value and to record this as incident.
- IDS should be capable of detecting duplicate options of Hop-by-Hop option header or destination options header.
- IDS with IPv6 support should also be able to recognize and analyze IPv6 traffic tunneled in IPv4.

Some of the existing works on intrusion detection in sensor network based IoT are mentioned below. Most of the researches are directed to wireless sensor network which is a main component of IoT. The lack of fixed infrastructures and scarce resource make WSN difficult to collect audit data for the entire network. It is more difficult to distinguish false alarms and real intrusions. In one of the previous papers [9], a survey of recent IDS in sensor network has been presented. In the earlier works, different lightweight hierarchical models [10][11] are proposed for heterogeneous wireless sensor network to detect DoS type attacks. But a sensor node can utilize most of the services offered by traditional IP networks with the help of IP stack. Without a proper security framework, it is not possible to grow towards IoT environment. So our research work currently focuses on this domain. In [20], a dynamic coding mechanism to implement distributed signature based lightweight IDS has been proposed in IP based Ubiquitous Sensor Networks. In [21], intrusion detection and response system has been proposed considering different types of attacks arise from internet hosts, clients and insider. Main module of the IDS resides on the gateway which supports dual stack- one for analyzing packets from Internet (IPA) and another for analyzing packets from USN (UPA) for detecting attacks. In [22], a distributed intrusion detection scheme for IoT has been proposed based on anomaly mining where intrusion semantic is analyzed to distinguish intrusion behaviors from anomalies; since all anomalies are not triggered by malicious intrusion. In [23], to detect the security threats in IoT, artificial immune system is applied where detectors evolve dynamically to detect new IoT attacks. Newly detected attacks are combined with the attack information library to alarm the manager of the IoT. The problem on how to consider new class of security threats in IP-enabled IoT is currently a challenging research issue.

Therefore a need of efficient security mechanism specifically tailored for this purpose is inevitable. In the next section, a pervasive environment based real life application of IoT has been briefly discussed to give idea about security implementation in next generation network.

## 6  Case Study: Pervasive Environment of Healthcare

A healthcare system based on pervasive environment controls health data in an electronic format EPR (Electronic Patient Record) as compared to the largely paper-based Records[24][25]. As EPRs are kept on networked systems for availability reasons, it is accessible from anywhere and is very easy to copy. Examples include confidential personal data like HIV status, psychiatric records, genetic information etc. The usage of EPRs, imposes new security risks to health data. This may lead to unauthorized access and tampering of sensitive EPRs. The pervasive use of wireless techniques makes it easy for malicious adversaries to launch security attacks. As healthcare systems are designed to assist in medical treatment, security vulnerabilities lead the entire system unreliable, putting patients' lives at risk. Some of the most possible vulnerabilities to this type of healthcare systems include the following.

- In case of emergencies, false alarms may be generated or real alarms may be suppressed by the system.
- Alteration of health data of specific patients, leading to incorrect diagnosis and treatment.

Another concern has been significantly increased over privacy issues, relating to electronic health data. Reliable healthcare systems must ensure same level of privacy policy for electronic data as applicable to paper based patient records. The main idea behind securing it is to preserve patient privacy. To ensure this, care needs to be taken to prevent all unauthorized access to EPRs in the system. An important property of a medical system is that patients have a high level of control over deciding who accesses their health information. Two additional issues associated with pervasive healthcare systems are security of wireless communication and physical security of handheld devices. Pervasive healthcare systems make extensive use of wireless communication technologies to communicate health data collected by sensor networks, which have many security vulnerabilities. Portable handheld devices that are used by both patients and caregivers, may store sensitive health information about the patient and cause a serious privacy breach, if stolen or misplaced. Therefore, physical security of the devices also has to be considered.

Recent development of cloud computing allows systems to store all or selective health data in cloud storage and ensures availability with reduced expenditures. In cloud design, data is stored on multiple third party servers where the storage can be accessed on demand. Migrating health data into the cloud offers enormous convenience to healthcare service providers because they do not have to worry about the complexities of direct hardware management. But user privacy preservation and proper access control of health data are growing concern. A contextual patient-centric access policy may be adopted to classify the roles. Security solutions of pervasive healthcare systems may focus on protecting health data from different aspects. In recent years several promising prototypes for wearable health monitoring have started to emerge. These devices are being used for continuous monitoring of patients for a long time. But effective security implementation for this type of pervasive applications is an unexplored area till date.

# 7   Conclusion

IoT is the emerging technology of the present age which consists of the analysis of new evolving data from heterogeneous sources for creating a new era of real life applica-tions. In this paper, an effort has been made to study the various security issues coming up in IoT environment. It is obvious that as IoT deals with a large number of objects, the use of wireless sensor network and cloud are almost inevitable. This gives rise to the need for transformation from IPv4 to IPv6 based connectivity to handle the huge num-ber of embedded smart objects. The use of IPv6 increases flexibility at the cost of vari-ous known and unknown security challenges. There have been many works on securing the wireless sensor networks already. The main focus in this paper is to study security threats in the new IPv6 enabled network, cloud environment and smart sensing "things" in the context of IoT. Without proper security framework, intelligence in IoT environ-ment may lead to major catastrophe. IoT research and innovation activities need to ad-dress security issues to support the growth of a smarter world. The possible solutions such as cryptography or intrusion detection etc. have been studied. The authors have considered a case study of healthcare system as this is definitely going to be maximum benefited in the IoT environment. This field is an emerging domain of research at present. The present survey is aimed at the construction of a secure independent living plan for the aged persons in the society in near future.